\documentclass{article}
\usepackage[margin=1in]{geometry}
\usepackage{jheppub}
\usepackage{physics}
\usepackage{enumitem}
\usepackage{amsmath,amssymb}
\usepackage[section]{placeins}
\usepackage{mathtools}
\usepackage{mdframed}
\usepackage{titlesec}
\usepackage{todonotes}
\usepackage{slashed}
\usepackage[normalem]{ulem}
\usepackage{tikz} 
\usepackage{tikz-feynman}
\tikzfeynmanset{compat=1.1.0}
\usepackage{chemformula}
\newcommand{\pd}{\partial}
\newcommand{\rmd}{\mathrm{d}}
\newcommand{\pdfr}[2]{\frac{\partial{#1} }{\partial{#2}}}
\newcommand{\dfr}[2]{\frac{\mathrm{d}{#1}}{\mathrm{d}{#2}}}
\newcommand{\He}{\ch{^4He}~}

\graphicspath{{figs/}}
\abstract{Ultralight dark matter is a compelling dark matter candidate. In this work, we examine the impact of quadratically-coupled ultralight dark matter on the predictions of Big Bang Nucleosynthesis. The presence of ultralight dark matter can modify the effective values of fundamental constants during Big Bang Nucleosynthesis, modifying the predicted abundances of the primordial elements such as Helium-4. We improve upon the existing literature in two ways: firstly, we take into account the thermal mass acquired by the ultralight dark matter due to its quadratic interactions with the Standard Model bath, which affects the cosmological evolution of the dark matter. Secondly, we treat the weak freeze-out using the full kinetic equations instead of using an instantaneous approximation. Both improvements were shown to impact the Helium-4 prediction in the context of universally-coupled dark matter in previous work. We extend these lessons to more general couplings. We show that with these modifications, Big Bang Nucleosynthesis provides strong constraints of ultralight dark matter with quadratic couplings to the Standard Model for a large range of masses as compared to other probes of this model, such as equivalence principle tests, atomic and nuclear clocks, as well as astrophysical and other cosmological probes.}

\title{Constraints on Ultralight Scalar Dark Matter with Quadratic Couplings}
\author[a,f]{Thomas Bouley,}
\author[b,c,d,e]{Philip Sørensen,}
\author[a,f]{and Tien-Tien Yu}
\emailAdd{tbouley@uoregon.edu}
\emailAdd{philip.soerensen@pd.infn.it}
\emailAdd{tientien@uoregon.edu}
\affiliation[a]{\small Department of Physics, University of Oregon,\\
Eugene OR 97403 USA
}
\affiliation[b]{\small Dipartimento di Fisica e Astronomia ‘G. Galilei’, Università di Padova,\\ Via F. Marzolo 8, 35131 Padova, Italy}
\affiliation[c]{\small INFN Sezione di Padova, Via F. Marzolo 8, 35131 Padova, Italy}
\affiliation[d]{\small II. Institute of Theoretical Physics, Universit\"{a}t  Hamburg, 22761 Hamburg, Germany}
\affiliation[e]{ \small Deutches Elektronen-Synchtron DESY, Notkestr. 85, 22607 Hamburg, Germany}
\affiliation[f]{\small  Institute for Fundamental Science, University of Oregon,\\
Eugene OR 97403 USA}

\begin{document}

\begin{flushright}
	\footnotesize
	DESY-22-186\\
\end{flushright}

\maketitle
\section{Introduction}
Most of the matter in the Universe is not accounted for in the Standard Model (SM) of particle physics. This missing matter is referred to as dark matter (DM). Proposed candidates for DM range from many solar mass compact objects to ultralight DM with masses as low as about $10^{-19}$ eV~\cite{Workman:2022ynf,Rogers:2020ltq,Dalal:2022rmp}.
In the low mass range, the DM is characterized by high occupation numbers, causing DM to behave like a classical field. 
Familiar examples of such ultralight DM (ULDM) are axion-like-particles~\cite{Preskill:1982cy,Abbott:1982af,Dine:1982ah} and dilatons (see, e.g., \cite{Damour:1994zq}). Dilatons are scalar fields $\phi$ present in models with compactified extra dimensions and have leading order couplings that are of the form \cite{Damour:2010rp}
\begin{equation}
	\frac{\phi}{\Lambda}\mathcal{O}_{\mathrm{SM}}\, ,
\end{equation} 
where $\mathcal{O}_{\mathrm{SM}}$ are terms in the SM Lagrangian and $\Lambda$ is the scale of new physics generating the coupling. Scalars with such couplings generate new long-range forces, and such theories are, therefore, heavily constrained by 5th force experiments \cite{EotWash_EP_2008,EotWash2020,HUST2021,IUPUI2016}. In this work, we will consider instead scalar ULDM with quadratic couplings of the form
\begin{equation}
	\frac{\phi^2}{\Lambda^2}\mathcal{O}_{\mathrm{SM}}\, ,
\end{equation} 
which evades the most stringent constraints on dilatons and possesses novel phenomenology distinct from models with linear couplings. Such quadratic couplings can arise if the new physics assumed at the high-energy (UV) scale $ \Lambda $ is subject to symmetries that restrict the linear contributions.
The lack of definitive observation of ULDM indicates that the ULDM must have extremely feeble couplings to ordinary matter.  
Beyond fifth-force searches, past attempts to search for ULDM have included searches for frequency variation of atomic clocks \cite{Clocks2015} and searches for perturbations in the arms of gravitational wave experiments \cite{ligo2021}.
Astrophysical probes of ULDM can be derived from Supernova SN1987A energy loss arguments~\cite{Raffelt:1987yt, Olive:2007aj}, from the spin-down of black holes due to superradiance~\cite{Arvanitaki:2010sy,Brito:2015oca,Cardoso:2018tly,Stott:2018opm,Stott:2020gjj,Mehta:2020kwu,Baryakhtar:2020gao}, and from pulsar timing arrays~\cite{Blas:2016ddr,Kaplan:2022lmz}.

In this work, we investigate cosmological probes of ULDM. Cosmology provides an attractive probe of ULDM because the field values of ULDM decrease with the expansion of the Universe, such that the largest field values are realized at early times. The earliest current probe of physics in the early Universe comes from Big Bang Nucleosynthesis (BBN), which therefore provides a powerful constraint of ULDM.
This constraint appears because the presence of a background DM field can shift the effective value of fundamental constants and therefore can modify BBN~\cite{Stadnik_2015}.
To adequately model the effect of ultralight DM on BBN, the backreaction from the SM bath on the ULDM must be accounted for. Furthermore, the BBN analysis must go beyond instantaneous approximations of the neutron-proton, or weak, freeze-out. Previous work~\cite{Sibiryakov_2020} has shown the relevance of these improvements in the context of universally-coupled ULDM. 
In this work, we investigate DM with more general couplings, correctly accounting for the evolution of the DM and the dynamics of weak freeze-out. 
We begin in Section~\ref{sec:model} with a description of the quadratically-coupled ULDM model and discuss the DM coupling to low-energy degrees of freedom. We also discuss the evolution of ULDM, properly accounting for the backreaction from the SM bath. In Section~\ref{sec:bbn}, we provide an overview of BBN and show how ULDM affects BBN, specifically the predictions of the Helium-4 (\He) abundance. In Section~\ref{sec:constraints} we will discuss the resulting constraints and compare the BBN constraints to other known constraints. We conclude in Section~\ref{sec:conclusions}.

\section{Ultralight Dark Matter With Quadratic Couplings}
\label{sec:model}
We begin by adding to the SM a real scalar field, 
\begin{equation}
	\mathcal{L}_\phi=\frac{1}{2}\pd_\mu\phi\pd^\mu\phi-\frac{1}{2}m^2_\phi\phi^2\,,
\end{equation}
where $m_\phi$ is the scalar mass. Couplings, such as the ones considered below, will lead to corrections to the bare scalar mass. Generically, these are of the form
\begin{align}
	\delta{m_\phi}&\approx d^{(2)}_f \frac{m_f^2 \Lambda_{\rm UV}^2}{M_{\rm pl}^2} \quad\text{coupling to fermions with mass $m_f$,}\\
	\delta{m_\phi}&\approx d^{(2)}_b \frac{\Lambda_{\rm UV}^4}{M_{\rm pl}^2}~\,\qquad\text{coupling to bosons,}
\end{align}
with $\Lambda_{\rm UV}$ the scale of the UV completion and $d^{(2)}_i$ are dimensionless couplings defined below. 
Na\"{i}vely, $ m_\phi $, smaller than this correction could be considered unnatural because this contribution would have to be tuned out. However, there exist models in which these contributions are suppressed, e.g., see \cite{Brzeminski:2020uhm,Banerjee:2022sqg}. Here, we seek to describe the phenomenology of such a field in a model-independent way, and we therefore leave any discussion of radiative corrections and protection from these to other work on model implementations.   

We consider scenarios in which the scalar field is charged under a symmetry, e.g., $\mathbb{Z}_2$, such that the leading coupling is quadratic, {\it i.e.}, we take the interactions of the scalar with the SM to be in the form $\frac{\phi^2}{\Lambda^2}\mathcal{O}_{\mathrm{SM}}$, where $\mathcal{O}_\mathrm{SM}$ is a term in the SM Lagrangian and $\Lambda$ is some high scale at which new physics appears. For ease of comparison with results in the literature, we will follow the conventions of~\cite{Damour:2010rm,Damour:2010rp,Hees_2018,Hees:2019nhn} and parameterize the interactions of the scalar particle with the SM with the following Lagrangian: 
\begin{equation}
	{\cal L}\supset 2\pi\frac{\phi^2}{M_{\rm pl}^2}\left[\frac{d_e^{(2)}}{4e^2}F_{\mu\nu}F^{\mu\nu}-\frac{d_g^{(2)} \beta_3}{2 g_3}G^A_{\mu\nu}G^{A\mu\nu}-d_{m_e}^{(2)}m_e\bar e e-\sum_{i=u,d}\left(d_{m_i}^{(2)}+\gamma_{m_i}d_g^{(2)}\right)m_i\bar\psi_i\psi_i\right]\, .
	\label{eq:Ldamour}
\end{equation}
Here, $M_{\rm pl}=1.22\times 10^{19}$ GeV is the Planck mass, $\beta_3$ is the QCD beta function, and $\gamma_{m_i}$ are the anomalous dimensions of the $u$ and $d$ quarks. The superscript $(2)$ specifies that these are the {\it quadratic} (as opposed to linear) couplings of the scalar.
An alternative convention for parameterizing these coupling found in \cite{Stadnik_2015,Stadnik:2014tta,Stadnik:2015xbn,Stadnik:2016zkf,Masia-Roig:2022net} has
\begin{equation}
	{\cal L}\supset \frac{\phi^2}{{\Lambda'}^2_e} \frac{1}{4e^2}F_{\mu\nu}F^{\mu\nu}-\frac{\phi^2}{{\Lambda'}^2_g} \frac{\beta_3}{2 g_3}G^A_{\mu\nu}G^{A\mu\nu}-\frac{\phi^2}{{\Lambda'}^2_{m_e}}m_e\bar e e-\sum_{i=u,d}\left(\frac{\phi^2}{{\Lambda'}^2_{m_i}}+\gamma_{m_i}\frac{\phi^2}{{\Lambda'}^2_g}\right)m_i\bar\psi_i\psi_i\, .
\end{equation}
This convention implies that the scale of the new physics mediating these couplings to the SM is around $\Lambda'$. The conversion from this convention to the one used in this work is given by $d^{(2)}_i=\frac{M_{\rm pl}^2}{2\pi {\Lambda'}^2_i}$ with $i=e,g, m_e, m_u, \mathrm{ or  } ~m_d$.

Note that the interactions defined in equation~\ref{eq:Ldamour} will give rise to $\phi$-dependency in the fundamental ``constants" of the SM. In a background $\phi$, the fundamental constants will shift according to    
\begin{align}
	\frac{\Delta\alpha}{\alpha}&=d_e^{(2)}\frac{\varphi^2}{2}\,,\label{eq:deltaalpha}\\
	\frac{\Delta\Lambda_{\rm QCD}}{\Lambda_{\rm QCD}}&=d_g^{(2)}\frac{\varphi^2}{2}\,,\label{eq:deltaQCD}\\
	\frac{\Delta m_f}{m_f}&=d_{m_f}^{(2)}\frac{\varphi^2}{2},~~~ \mathrm{for } ~f=e,u,d,~~\label{eq:deltame}
\end{align}      
where we have defined $\varphi=\frac{\sqrt{4\pi}\phi}{M_{\rm pl}}$.  Here, $\alpha\simeq 1/137$ is the fine-structure constant, $\Lambda_{\rm QCD}$ the QCD confinement scale, and $m_f$ are the fermion masses.
The quark mass-couplings are more useful written in terms of the symmetric and antisymmetric combinations:
\begin{align}
	d_{\hat m}^{(2)}&\equiv \frac{d_{m_d}m_d+d_{m_u}m_u}{m_d+m_u}~~~~~~~\text{symmetric,}\\
	d_{\delta m}^{(2)}&\equiv \frac{d_{m_d}m_d-d_{m_u}m_u}{m_d-m_u}~~~~~~\text{anti-symmetric,}
\end{align}
as physical quantities come in these combinations. In particular, the quark mass-couplings enter BBN through the neutron-proton mass difference, which depends on the anti-symmetric combination, and through the neutron axial coupling, which depends on the symmetric combination. 

\subsection{Dark matter evolution}\label{sec:DMevo}
\begin{figure}[h]
	\centering
	\includegraphics[width=13cm]{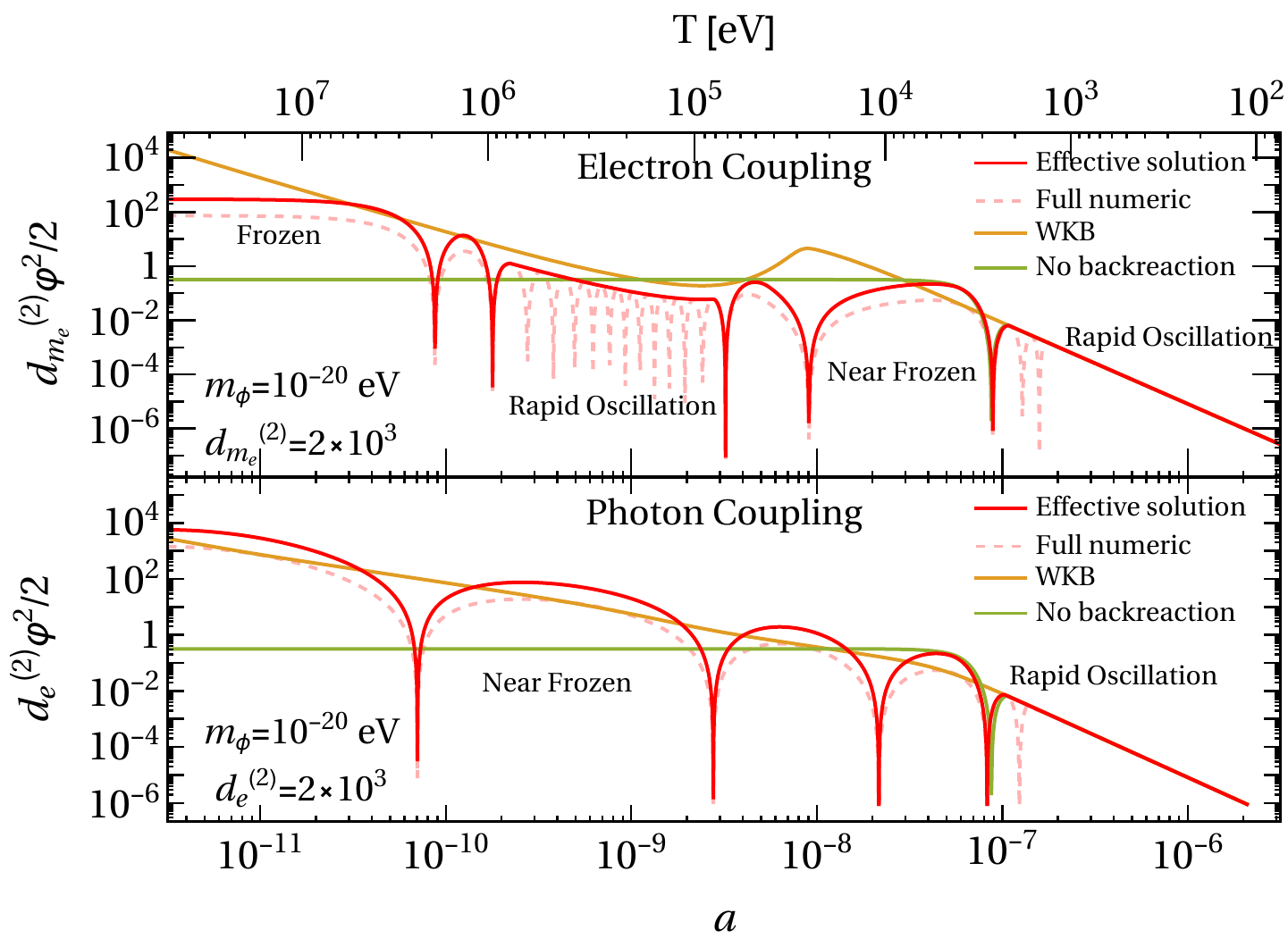}
	\caption{The evolution of the DM coupled to electrons (upper) or photons (lower). The effective solution is a splice of the numerical solution and the WKB approximation. The thermal mass used for the photon evolution was computed in the high-temperature limit as explained in section~\ref{sec:photonmass}.   The difference between the numerical solution and the effective solution in the non-oscillating regions is due to factors of 2 present in averaging. Time is indicated both in terms of temperature $ T $ and in terms of the scale factor $ a $, the latter of which is normalized to $ a_0=1 $ today.}
	\label{fig:evolutionexample}
\end{figure}
We are interested in the field values of the ULDM at BBN. The present-day field value, $\phi_0$, is fixed by the current abundance of DM: $\rho_{{\rm DM},0}=\frac{1}{2}m_\phi^2\phi_0^2$. To determine the value of the field around BBN, we evolve the scalar field backward in time according to its equations of motion, using the present-day field value $\phi_0$ as a boundary condition,
\begin{equation}
	\ddot{\phi}+3 H \dot{\phi}+m^2_{\mathrm{eff}}\phi=0 \, ,
	\label{eq:EOM}
\end{equation} 
where the $\dot~$ denotes derivatives with respect to time. The effective mass is a combination of the bare mass and the thermal mass induced by the SM bath: $m_{\rm eff}^2=m_\phi^2+m_{\rm ind}^2$. During BBN, the SM bath is primarily composed of photons and electrons. Thus, only interactions with photons ($d_e^{(2)}$) and electrons ($d_{m_e}^{(2)}$) give significant contributions to the thermal mass as other species are not sufficiently abundant. Therefore, the quark ($d_{\delta m}^{(2)}$ and $d_{\hat m}^{(2)}$) and gluon ($d_g^{(2)}$) couplings do not provide a significant contribution to the thermal mass and the backreaction from these two couplings can be ignored.

To solve equation~\ref{eq:EOM}, we can examine three distinct regimes of DM evolution:
\begin{itemize}
	\item Hubble friction domination (H): When $H^2\gg m_{\mathrm{eff}}^2$
	\item Bare mass domination (B): When $m_\phi^2\sim m^2_{\mathrm{eff}}\gg H^2$
	\item Induced mass domination (I): When $m_{\mathrm{eff}}^2\gg H^2, m_\phi^2$ 
\end{itemize}
In the B and I regimes, $\phi$ is rapidly oscillating, and performing the full computation is numerically expensive. Instead, we model the interactions of the ULDM with matter by taking the average over an oscillation period. We use a WKB-type solution, as in \cite{Sibiryakov_2020}, to find the amplitude of the oscillation. This WKB-type approximation gives,
\begin{equation}
	\phi_{\mathrm{amp}}\propto m_{\mathrm{eff}}^{-1/2} a^{-3/2},
\end{equation}
where $\phi_\mathrm{amp}$ is the amplitude of the rapidly oscillating  $\phi$.
In the H regime, the field value is constant due to the Hubble friction term dominating the equation of motion. In the transitions between the regimes, the field is slowly oscillating and we can solve the field evolution numerically and resolve the behavior of $\phi$ without the need for averaging.  Figure \ref{fig:evolutionexample} shows an example evolution for $m_\phi=10^{-20}\mathrm{eV}$ and $d_{e}^{(2)}=2000$ or $d_{m_e}^{(2)}=2000$ comparing the evolution computed using our method to the pure WKB and numerical solution. In the top panel showing the electron coupling case the field transitions from being Hubble dominated at early times, to being induced mass dominated, and finally to being bare mass dominated. In the induced mass domination regime the field is slowly oscillating when the induced mass is comparable to Hubble. We see the same behavior in the bottom panel, which shows the photon coupling case. We see that at early times the induced mass is comparable to Hubble and the field is slowly oscillating, and at late times it transitions to bare mass domination. In both panels, we see that at late times the field is rapidly oscillating, which is needed to satisfy the requirement that the field is DM today. The behavior shown in figure \ref{fig:evolutionexample} is not generic and the evolution depends on the mass and couplings.
As we see the induced mass is crucial to understand the behavior of the scalar field. In what follows, we discuss the induced thermal mass for the electron and photon couplings. 

\subsubsection{Electron coupling}
\begin{figure}
	\centering
	\includegraphics[width=13cm]{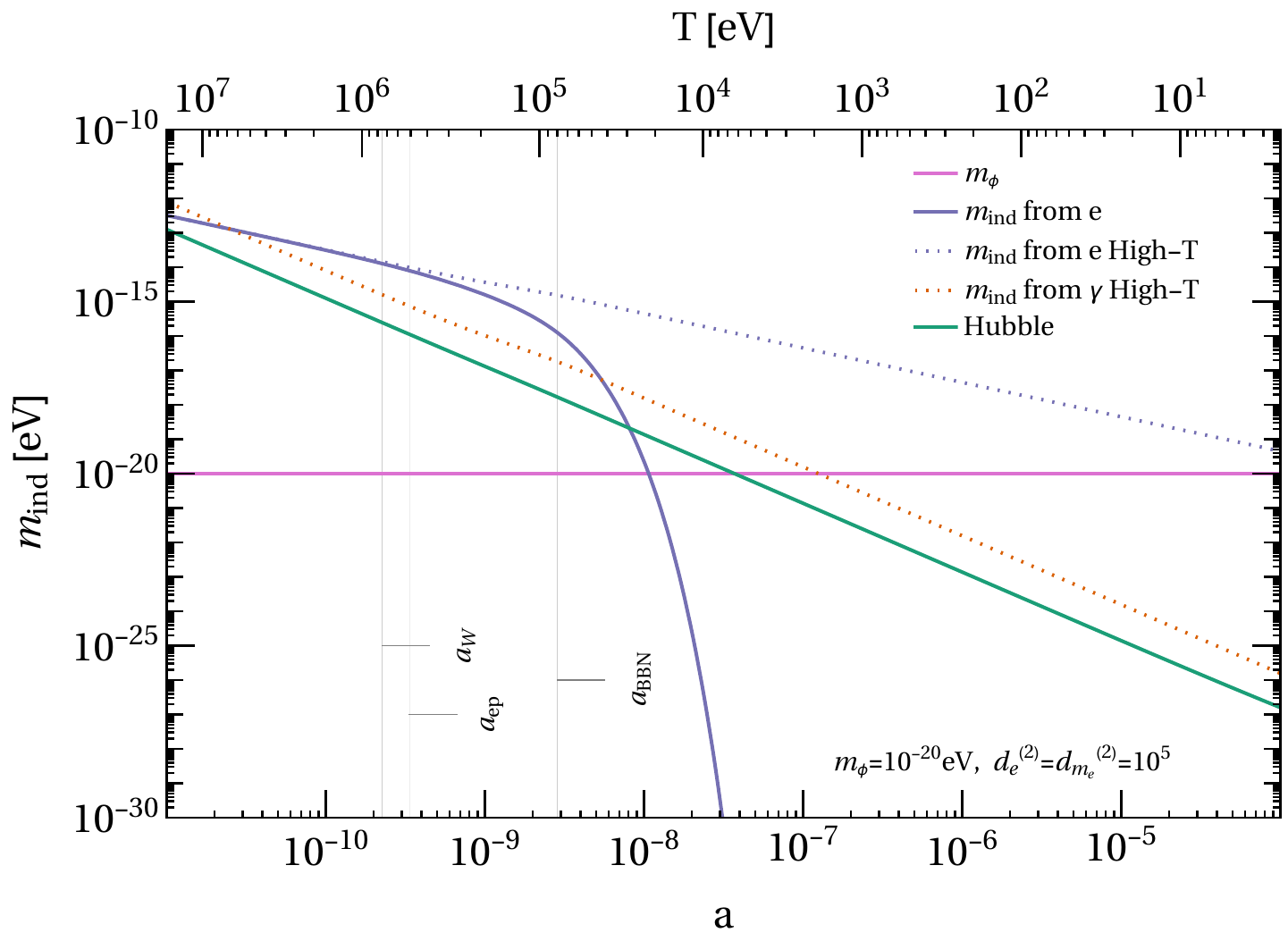}
	\caption{Potential contributions to $m_\mathrm{eff}$ from the electron and photon coupling for the couplings  $d_{e}^{(2)}=10^5$ or $d_{m_e}^{(2)}=10^5$ . The dotted lines are the high-temperature approximations for the electron and photon couplings, respectively.} 
	\label{fig:contributions}
\end{figure}
There are two ways to calculate the induced mass from the interactions with electrons.
The first is to calculate the induced mass using thermal field theory. The leading diagram is\\
\begin{equation}
	\begin{tikzpicture}[baseline=(current bounding box.center)] 
		\begin{feynman}
			\vertex (a) {$\phi$};
			\vertex [above right=1cm and 1cm of a] (c);
			\vertex [right=2cm of a] (b) {$\phi$};
			
			\diagram* {
				(a) -- (c) --(b),
				(c) --[fermion, edge label=$e$, loop,min distance=2cm, out=135, in=45] c
			};
		\end{feynman}
	\end{tikzpicture} \, .
\end{equation}
In the imaginary time formalism described in~\cite{KAPUSTA1979s, Quiros:1999jp},
this diagram gives
\begin{equation}\label{eq:mePi1}
	\Pi_1=-2 \frac{2 \pi d^{(2)}_{m_e}m_e}{M_{\rm pl}^2} T\sum_{n=-\infty}^{\infty}\int\frac{\rmd^3 p}{(2\pi)^3}\frac{\mathrm{tr}(\slashed{p}+m_e)}{p^2-m_e^2}\,,
\end{equation} 
where the sum is over the $n$ Matsubara frequencies which each fix the time component of $p$ to  $p_0 = i(2n+1)T$.
The temperature-dependent part of this diagram is given by 
\begin{equation}\label{eq:meffe}
	m_{\mathrm{eff}}^2=m_\phi^2+\frac{2\pi d^{(2)}_{m_e}}{M_{\rm pl}^2}\frac{4 m_e^2}{\pi^2} T^2\int^\infty_{m_e/T}\rmd x \frac{\sqrt{x^2- (m_e/T)^2}}{e^x+1}\, .
\end{equation} 
This calculation can be cross-checked with a second way of calculating the effective mass, which is to note that when electrons are on-shell the interaction $\frac{2\pi d_{m_e}^{(2)}}{M_{\rm pl}^2}\phi^2m_e\bar{\psi}\psi$ can be written as $\frac{2\pi d_{m_e}^{(2)}}{M_{\rm pl}^2}\phi^2\frac{i}{2}(\bar{\psi}\slashed{\pd}\psi-(\pd_\mu\bar{\psi})\gamma^\mu\psi)=\frac{2\pi d_{m_e}^{(2)}}{M_{\rm pl}^2}\phi^2\Theta_e$ where $\Theta_e$ is the trace of the electrons' contribution to the stress-energy tensor. The effective mass is then given by \cite{Sibiryakov_2020,Erickcek_2014}
\begin{equation}
	m_{\mathrm{eff}}^2=m_\phi^2+\frac{4\pi d_{m_e}^{(2)}}{M_{\rm pl}^2}{\Theta_e}=m_\phi^2+\frac{4\pi d_{m_e}^{(2)}}{M_{\rm pl}^2}(\rho_e-3P_e)\, ,
\end{equation} 
which agrees with equation \ref{eq:meffe} after substituting
\begin{align}
	\rho_e&=\frac{2}{\pi^2}T^4\int^\infty_{m_e/T} \rmd x \frac{x^2\sqrt{x^2- (m_e/T)^2}}{e^x+1},\\
	P_e&=\frac{2}{3\pi^2}T^4\int^\infty_{m_e/T} \rmd x \frac{(x^2- (m_e/T)^2)^{3/2}}{e^x+1}.
\end{align}

Note that for $T\gg m_e$, equation \ref{eq:meffe} implies that $m_{\mathrm{eff}}^2\propto T^2$. In this regime, $m_{\mathrm{eff}}^2 \propto a^{-2}$ whereas $H^2\propto a^{-4}$. Thus, at early times (high temperature), the DM evolution is dominated by Hubble friction and the field is in the H regime. 
Thus, the mass induced on $\phi$ from finite density effects can lead to highly non-trivial evolution of the scalar field. This can be seen in figure \ref{fig:contributions}, where the induced mass $m_{\rm ind}^{2}\equiv m_{\rm eff}^2-m_\phi^2$ is compared to both $H$ and the bare mass $m_\phi$. In the example given in figure \ref{fig:contributions}, which corresponds to the behavior also seen in figure \ref{fig:evolutionexample},  the field is initially dominated by Hubble friction, then starts oscillating because of the induced mass and the returns to a brief Hubble friction dominated state before finally settling into an oscillating state driven by the late time mass. Figure \ref{fig:contributions} also shows the mass that can be induced by a photon coupling, which are discussed in the next section.

\subsubsection{Photon coupling}\label{sec:photonmass} 
For the photon coupling, we calculate the contribution to the induced mass using thermal field theory. The 1-loop diagram,  
\begin{equation}
	\begin{tikzpicture}[baseline=(current bounding box.center)] 
		\begin{feynman}
			\vertex (a) {$\phi$};
			\vertex [above right=1cm and 1cm of a] (c);
			\vertex [right=2cm of a] (b) {$\phi$};
			
			\diagram* {
				(a) -- (c) --(b),
				(c) --[photon, edge label=$\gamma$, loop,min distance=2cm, out=135, in=45] c
			};
		\end{feynman}
	\end{tikzpicture} \, ,
\end{equation}
gives
\begin{equation}
	\Pi_1=\frac{2\pi d_{m_e}^{(2)}}{ M_{\rm pl}^2} T\sum_{n=-\infty}^{\infty}\int\frac{\rmd^3 p}{(2\pi)^3}(g^{\mu\nu}p^2-p^\nu p^\mu)\frac{g_{\mu\nu}}{p^2}\, ,
\end{equation}
where similarly to equation \ref{eq:mePi1} the sum is over the $n$ Matsubara frequencies which set $p_0=i 2n\pi T$. This gives a scaleless integral and so vanishes in dimensional regularization, therefore this diagram does not contribute to the thermal mass. Instead, the leading diagram appears at two loops 
\begin{equation}\label{feyn:photon_leading}
	\begin{tikzpicture}[baseline=(current bounding box.center)] 
		\begin{feynman}
			\vertex (a) {$\phi$};
			\vertex [above right=1.5cm and 1.5cm of a] (c);
			\vertex [right=3cm of a] (b) {$\phi$};
			\vertex [above left=1.5cm and .5cm of c](d);
			\vertex [above right=1.5cm and .5cm of c](e);
			
			\diagram* {
				(a) -- (c) --(b),
				(c) --[photon, quarter left] (d),
				(e) --[photon, quarter left] (c),
				(d) --[fermion, half left,edge label=$e$] (e),
				(e) --[fermion, half left,edge label=$e$] (d)
			};
		\end{feynman}
	\end{tikzpicture} \, .
\end{equation}
This gives
\begin{equation}
	\Pi_2=\frac{2 \pi d^{(2)}_e}{M_{\rm pl}} e^2 T\sum_{n=-\infty}^{\infty}\int\frac{\rmd^3 k}{(2\pi)^3}T\sum_{m=-\infty}^{\infty}\int\frac{\rmd^3 p}{(2\pi)^3}\frac{(g^{\mu\nu}k^2-k^\mu k^\nu)(-1)\tr(\gamma_\mu(\slashed{p}+m_e)\gamma_\nu(\slashed{p}-\slashed{k}+m_e))}{k^4 (p^2-m_e^2)((p-k)^2-m_e^2)}\,,
\end{equation}
summing over the $n$ and $m$ Matsubara frequencies which set  $k_0=i 2n\pi T$ and $p_0=i (2m+1)\pi T$ . Evaluating the above expression is substantially complicated due to overlapping UV divergences, but it is relatively simple to compute in the limit where $T\gg m_e$. In this high-temperature limit, the contribution to the thermal mass is given by
\begin{equation}
	\Pi_2=\frac{2 \pi d^{(2)}_e}{M_{\rm pl}^2}\frac{\alpha}{4\pi}\frac{\pi^2}{3}T^4\,.
	\label{eq:photonHighT}
\end{equation}
This high-temperature limit is a good approximation for the thermal mass at temperatures much larger than the electron mass.  For lower temperatures, the finite electron mass suppresses the contribution. Therefore, we expect the high-temperature result to overestimate the thermal mass.

In the regime where the field is always oscillating, we expect this thermal mass contribution to somewhat relax the constraints. Therefore, the high-temperature result can be taken as conservative. This can be seen from the WKB-approximation, by which the amplitude of the oscillations can be described as $\phi\propto m_{\rm eff}^{-1/2} a^{-3/2}$.
The thermal mass contribution decays in time, which then tends to compensate for the growth in $ a $ and therefore slow down the redshift of $ \phi $. This reduces the hierarchy between $ \phi_{\rm BBN} $ and $ \phi_0 $. Since $ \phi_0 $ is fixed by the zero-temperature mass and the observed DM abundance today $ \phi_{\rm BBN} $ will therefore be smaller in a scenario with thermal mass contribution, as long as WKB is always a good approximation of the evolution. This argument breaks down in the regime in which Hubble friction becomes significant and WKB approximation fails. In this regime, a thermal mass contribution can overcome Hubble friction which increases the strength of the constraint. At low $ m_\phi $ the impact on the constraint is therefore a competition between opposing effects. For our purposes, we will present our results in the two limits using the high-temperature effective mass given in equation~\ref{eq:photonHighT} and in the low-temperature limit where $m_{\rm eff}\simeq m_\phi$, which provides an envelope within which the true bound will sit. 

\section{Ultralight Dark Matter and BBN} 
\label{sec:bbn}
We are interested in understanding the effect of the ULDM couplings on the predictions of BBN. We, in section \ref{sec:bbn review}, begin with a brief review of the standard BBN analysis, following the discussion in \cite{Mukhanov_2004}. From this, we can calculate an analytic estimate of the \He abundance. In section \ref{sec:bbnEffect}, we perturb the standard result to obtain constraints.

\subsection{Standard BBN} 
\label{sec:bbn review}
The \He abundance is determined by the abundance of neutrons at the time that fusion to helium becomes efficient. This in turn is determined by the abundance of neutrons when neutrinos decouple and the time at which deuterium can form. Before weak freeze-out, the weak reactions $n+e^+\leftrightarrow p+\bar{\nu}_e$ and $n+\nu_e\leftrightarrow p+e^-$ are efficient and neutrons and protons have equilibrium abundances, 
\begin{align}
	X_n^{eq}&=\frac{1}{1+e^{m_{np}/T}}\,,\\
	X_p^{eq}&=1-X_n^{eq}=\frac{1}{1+e^{-m_{np}/T}}\,,
\end{align}
where $X_n^{eq}$ and $X_p^{eq}$ are the equilibrium fractions of baryons in neutrons and protons, respectively, and $m_{np}$ is the proton-neutron mass difference.  
As the Universe expands and cools, the reaction rate for these processes falls below the expansion rate of the Universe, the Hubble scale, and so the neutron abundance falls out of equilibrium. At this point, the evolution of the neutron abundance $X_n$ is governed by the kinetic equation, 
\begin{equation}\label{eq:diff}
	a\dfr{X_{n}}{a}=-\frac{\lambda_{n\to p}}{H}(1+e^{-m_{np}/T})(X_n-X_n^{eq})\, ,
\end{equation}
where $a$ is the scale factor (normalized to $ a_0=1 $ today), $H=\dot a/a$ is the Hubble parameter, $T$ is the temperature, and $\lambda_{n\to p}$ is the reaction rate of $n+(e^+\ \mathrm{or}\  \nu)\to p +(\bar{\nu}\ \mathrm{or}\ e^-)$: 
\begin{equation}
	\lambda_{n\to p}=\frac{1+3 {g_A}_n}{\pi^3}G_F^2 T^5 J\left(\frac{m_{np}}{T}\right)\, .
\end{equation}
Here, $G_F$ is the Fermi constant, $g_{A_n}$ is the weak axial coupling, and $J$ is a phase space factor, given by
\begin{equation}
	J(z)=\frac{45\zeta(5)}{2}+\frac{21\zeta(4)}{2} z+\frac{3\zeta(3)}{2}\left(1-\frac{m_e^2}{2m_{np}^2}\right) z^2\, .
\end{equation}
Solving equation \ref{eq:diff} for the neutron abundance at late times gives
\begin{equation}\label{eq:XnW}
	X_{n,W}=-\int^\infty_0 \rmd a \dfr{X^{eq}_n}{a} \mathrm{exp} \bigg[-\int^\infty_a \frac{\rmd a'}{a'} \tilde{\lambda}_{n\to p} \bigg]\, ,
\end{equation} 
with $ \tilde{\lambda}_{n\to p}=\frac{\lambda_{n\to p}}{H}(1+e^{-m_{np}/T})$.\\

At this point, the neutron abundance continues to decrease due to neutron decay.  The neutron abundance at these times follows
\begin{equation}\label{eq:decay_diff}
	a \dfr{X_n}{a}=-\frac{X_n \Gamma_n}{H}\, ,
\end{equation}   
with $\Gamma_n$ the neutron inverse lifetime given by
\begin{equation}
	\Gamma_n=\frac{1+3 {g_A}_n^2}{2 \pi^3}G_F^2 m_e^5 P\left(\frac{m_{np}}{m_e}\right)\, ,
\end{equation}
where $P$ is a phase space factor given by
\begin{equation}
	P(x)=\frac{1}{60}\left((2x^4-9x^2-8)\sqrt{x^2-1}+15x\ln{(x+\sqrt{x^2-1})}\right)\, .
\end{equation}
Integrating equation \ref{eq:decay_diff}, we find the neutron abundance at BBN:
\begin{equation}\label{eq:xbbn}
	X_{n,BBN}=X_{n,W}\mathrm{exp}\bigg(-\int^{a_{BBN}}_{a_W}\frac{\rmd a}{a H}\Gamma_n\bigg)\, .
\end{equation}

After weak freeze-out, neutrons and protons fuse to form deuterium through 
\begin{equation}
	p+n\leftrightarrow D+\gamma\, .
	\label{eq:deuterium}
\end{equation}
Initially, the photon abundance is much larger than the baryon abundance, and given that the deuterium binding energy is not very large compared to the temperature, equilibrium strongly favors the left side of equation~\ref{eq:deuterium}. The deuterium abundance is given by the Saha equation
\begin{equation}\label{eq:saha}
	X_D=\frac{24\zeta(3)}{\sqrt{\pi}}\eta_b X_p X_n \left(\frac{T}{m_p}\right)^{3/2}e^{B_D/T}\, ,
\end{equation} 
where $X_{p,n}$ is the abundance of protons and neutrons, respectively, $B_D$ is the deuterium binding energy, and $\zeta$ is the Riemann zeta function. Note that the \He binding energy is much larger than $B_D$ so in equilibrium \He would dominate. However, the reactions needed to form helium
\begin{align}
	\ch{D + D&<-> T + p}\, ,\nonumber\\
	\ch{D + D&<-> ^3He + n}\, ,\label{eq:Heformation}\\
	\ch{D + T&<-> ^4He + n}\, ,\nonumber\\
	\ch{D + ^3He&<-> ^4He + p}\, ,\nonumber
\end{align}
depend on the formation of deuterium and are suppressed by the low abundance of deuterium. This suppression of the deuterium abundance due to the background photons is known as the ``deuterium bottleneck". Eventually when $T\approx\frac{B_D}{30}$, the deuterium abundance becomes large enough such that the reactions \ref{eq:Heformation} become efficient. At this time most of the neutrons become bound into helium and the \He abundance is
\begin{equation}\label{eq:XHe}
	Y_p\approx 2X_{n,BBN}\, .
\end{equation} 
Not long after this, the Universe becomes too cool and diffuse for these nuclear reactions to be efficient. This signals the end of BBN. After this, but before the onset of star formation, the tritium and \ch{^7Be} decay to \ch{^3He} and \ch{^7Li} respectively\footnote{It should be noted that in the standard theory there was a discrepancy between predictions and observations of the \ch{^7Li} abundance. This is known as the \textit{lithium problem}. However, the existence of this tension has recently been put into question~\cite{Fields:2022mpw}.}. 

Equation \ref{eq:XHe} gives $Y_p\approx0.25$ \cite{Mukhanov_2004}. A detailed numerical calculation including the kinetics of the fusion reactions given in \ref{eq:Heformation}  gives $Y^{th}_p=0.24672\pm0.00061$ using the baryon to photon ratio measured by Planck~\cite{Planck:2018vyg}. This is in excellent agreement with the PDG recommended average of $Y^{ex}_p=0.245\pm0.003$~\cite{Workman:2022ynf}. 

\subsection{Effect of varying fundamental constants} 
\label{sec:bbnEffect}
We now consider how the presence of ULDM modifies the standard analysis above. We first consider how $ \phi $ varies the fundamental constants and we then consider how this impacts the neutron fraction at freeze-out and at BBN. Finally, we derive the impact of $ \phi $ on \He.

\subsubsection{Variation in fundamental constants relevant for BBN      }
\label{sec:BBNvar}
As we saw in the previous section, the \He abundance produced by BBN depends on the deuterium binding energy $B_D$, the proton-neutron mass difference $m_{np}$, and the neutron axial coupling $g_A$. In order to determine how the presence of $\phi$ affects BNN, one must determine how $\phi$ shifts $m_{np}$,  $B_D$, and $g_A$.  These parameters do not have analytic expressions but can only be calculated from first principles using lattice gauge theory. 
The shift in $m_{np}$ and $B_D$ can be estimated using a combination of lattice and analytic methods as discussed in~\cite{Coc_2007};  we summarize the relevant parts below.
The neutron-proton mass difference can be written as
\begin{equation}
	m_{np}=m_n-m_p\approx b\alpha \Lambda_\mathrm{QCD} +(m_d-m_u)\, ,
\end{equation} 
where $b$ is some constant number determined as in \cite{Coc_2007}, giving $b\alpha \Lambda_\mathrm{QCD}\approx-0.76\mathrm{MeV}$.
Using equations \ref{eq:deltaalpha}, \ref{eq:deltaQCD}, and \ref{eq:deltame} we get 
\begin{equation}\label{eq:deltamnp}
	\frac{\Delta m_{np}}{m_{np}}={\frac{b\alpha \Lambda_\mathrm{QCD}}{m_{np}}\left({d_e^{(2)}}+{d_g^{(2)}}\right)\frac{\varphi^2}{2}+\frac{m_d-m_u}{m_{np}}\left({d_{\delta m}^{(2)}+\gamma d_g^{(2)}}\right)\frac{\varphi^2}{2}}\, .
\end{equation}
As already stated above, the dimensionless parameter $ \varphi $ is defined as $ \varphi=\frac{\sqrt{4\pi}\phi}{M_{\rm pl}} $. The dependence of $B_D$ on the fundamental constants can be parameterized as \cite{Coc_2007}
\begin{equation}\label{eq:deltaBD}
	\frac{\Delta B_D}{B_D}=18\frac{\Delta \Lambda_\mathrm{QCD}}{\Lambda_\mathrm{QCD}}=18{d_g^{(2)}}\frac{\varphi^2}{2} \, .
\end{equation}

Determining the dependence of $g_A$ on fundamental constants is more involved and requires input from lattice results. Lattice-QCD based attempts to estimate $g_A$ from first principles have limited statistics for physical quark masses and rely on calculations at heavier quark masses to fit formulae taken from chiral perturbation theory (ChPT) to extrapolate to physical quark masses. To estimate the dependence of $g_A$ on the quark masses we use these fits. Specifically, we use the expression 
\begin{equation}\label{eq:g_Aofe}
	g_A(\epsilon_\pi)=g_0-\epsilon_\pi^2\left((g_0+2g_0^3)\ln{\epsilon_\pi}-c_2\right )+g_0 c_3 \epsilon_\pi^3
\end{equation}
with $\epsilon_\pi=\frac{m_\pi}{4\pi f_\pi}$. This expression is from NNLO heavy baryon ChPT. It neglects effects from $\Delta$ resonances \cite{Hemmert:2003cb,Kambor:1998pi} and is used by lattice groups for their extrapolation to physical quark masses \cite{Chang:2017oll,Walker-Loud:2019cif}. The coefficients $g_0$, $c_2$, and $c_3$ are determined from fits to lattice calculations. The dependence of $g_A$ on $\epsilon_\pi$ can be used to infer the dependence on the quark masses using $m_\pi\propto \sqrt{\hat{m}}$ and neglecting the small dependence of $f_\pi$ on the quark masses. Putting this all together, we have
\begin{equation}
	\frac{\Delta g_A}{g_A}=d_{\hat m}^{(2)}\varphi\frac{1}{2} \pdfr{\ln g_A}{\ln \epsilon_\pi}
\end{equation}
This depends on the specific values for the coefficients $g_0$, $c_2$, and $c_3$ that come from lattice fits. The values obtained from \cite{Chang:2018uxx} and its supplementary materials are
\begin{align}
	g_0&=1.237\pm 0.034\, ,\\
	c_2&=-23.0\pm 3.5\, ,\\
	c_3&=28.7\pm 5.5 \, ,
\end{align}
with the covariance matrix 
\begin{equation}
	\begin{array}{c | c c c}
		& g_0 & c_2 & c_3 \\ \hline
		g_0   & 0.011 & -0.12 & 0.18 \\
		c_2   & -0.12 & 12.25 & -19.17\\
		c_3 	& 0.18 & -19.17 & 30.25 
	\end{array}\, .
\end{equation}
These give
\begin{equation} 
	\pdfr{\ln g_A}{\ln \epsilon_\pi}=-0.008\pm 0.023 \, .
\end{equation}
Since the dependence of the \He abundance is proportional to $d^{(2)}_{\hat{m}}\pdfr{\ln g_A}{\ln \epsilon_\pi}$, and the interval above contains zero, we are not able to give meaningful constraints on $d^{(2)}_{\hat{m}}$ from BBN at this time.
Improved lattice results may constrain $\pdfr{\ln g_A}{\ln \epsilon_\pi}$ away from 0, which in turn could give constraints on $d_{\hat{m}}^{(2)}$ from BBN. 

Armed with these expressions that encode the effects of the ULDM on the parameters that enter the calculation for the \He abundance, we can now determine the effect of ULDM on the \He abundance.
\subsubsection{Variation in the neutron abundance after weak freeze-out}
We begin by varying \ref{eq:XnW} to find the change in $X_{n,W}$ due to the presence of $\phi$ (or equivalently to $\varphi$). To first order in $\kappa=d^{(2)}_{i}\frac{\varphi^2}{2}$ , where $i=e,m_e,g,~\mathrm{ or }~ \delta m$, we find
\begin{align}
	\Delta X_{n,W}&=-\int^\infty_0 \rmd a\Bigg( \dfr{X^{eq}_n}{a} \mathrm{exp} \bigg[-\int^\infty_a \frac{\rmd a'}{a'} \tilde{\lambda}_{n\to p} \bigg]\bigg(-\int^\infty_a\frac{\rmd a'}{a'}{\Delta\tilde{\lambda}_{n\to p}}\bigg)\nonumber\\
	&\qquad\qquad +\dfr{\Delta X^{eq}_n}{a} \mathrm{exp} \bigg[-\int^\infty_a \frac{\rmd a'}{a'} \tilde{\lambda}_{n\to p} \bigg]  \Bigg)\\
	&=-\int^\infty_0 \rmd a\Bigg( \dfr{X^{eq}_n}{a} \mathrm{exp} \bigg[-\int^\infty_a \frac{\rmd a'}{a'} \tilde{\lambda}_{n\to p} \bigg]\bigg(-\int^\infty_a\frac{\rmd a'}{a'}{\Delta\tilde{\lambda}_{n\to p}}\bigg)\nonumber\\
	&\qquad\qquad -\Delta X^{eq}_n \frac{\tilde{\lambda}_{n\to p}}{a}\mathrm{exp} \bigg[-\int^\infty_a \frac{\rmd a'}{a'} \tilde{\lambda}_{n\to p} \bigg]  \Bigg)\\
	&=-\int^\infty_0 \frac{\rmd a}{a}\mathrm{exp} \bigg[-\int^\infty_a \frac{\rmd a'}{a'} \tilde{\lambda}_{n\to p} \bigg] \bigg(-\Delta X^{eq}_n \tilde{\lambda}_{n\to p}-a \dfr{X^{eq}_n}{a} \int^\infty_a\frac{\rmd a'}{a'}{\Delta\tilde{\lambda}_{n\to p}}\bigg)\,.
\end{align}
The change in $\Delta X_{n,W}$ depends on $\Delta X^{eq}_n$, $\dfr{X^{eq}_n}{a}$, and $\Delta \tilde{\lambda}_{n\to p}$. Let's first examine the variation in $X^{eq}_n$
\begin{equation}
	\Delta X^{eq}_n=-\frac{m_{np}}{2T(1+\cosh{m_{np}/T})}\frac{\Delta m_{np}}{m_{np}}\, ,
\end{equation}
and compare this to
\begin{equation}
	a\dfr{X^{eq}_n}{a}=-\frac{m_{np}}{2 T(1+\cosh(m_{np}/T))}\, .
\end{equation} 
The variation in $\tilde{\lambda}_{n\to p}$ is given by
\begin{equation}
	\frac{\Delta \tilde{\lambda}_{n\to p}}{\tilde{\lambda}_{n\to p}}=\frac{\Delta \lambda_{n\to p}}{\lambda_{n\to p}}-\frac{m_{np}X^{eq}_n}{T}\frac{\Delta m_{np}}{m_{np}}\, ,
\end{equation}
with the variation in $ \lambda_{n\to p}$ given by
\begin{align}
	\frac{\Delta \lambda_{n\to p}}{\lambda_{n\to p}}&=\frac{6 {g_A}_n^2}{1+3{g_A}_n^2}\frac{\Delta {g_A}_n}{{g_A}_n} +2\frac{\Delta G_F}{G_F} +\frac{\Delta (J(m_{np}/T))}{J(m_{np}/T)} \\
	&=\frac{6 {g_A}_n^2}{1+3{g_A}_n^2}\frac{\Delta {g_A}_n}{{g_A}_n}+2\frac{\Delta G_F}{G_F} +\frac{m_{np} J'}{TJ}\frac{\Delta m_{np}}{m_{np}}\nonumber\\
	&\qquad\qquad-\frac{3\zeta(3)}{2J}\frac{m^2_e}{T^2}\frac{\Delta m_{e}}{m_{e}}+\frac{3\zeta(3)}{2J}\frac{m^2_e}{T^2}\frac{\Delta m_{np}}{m_{np}}\, . \label{eq:Dlamdba}
\end{align}
The last two terms in equation~\ref{eq:Dlamdba} are proportional to $\frac{m_e^2}{T^2}$, which is small at relevant times and in principle can be neglected. However, we will include them going forward. Note that $\frac{\Delta G_F}{G_F}=0$ for the models we are considering.
Putting everything together we find that the change in the neutron abundance at weak freeze-out is given by
\begin{mdframed}
	\begin{equation}
		\begin{aligned}
			\Delta X_{n,W}=\int^\infty_0& \frac{\rmd a}{a}\frac{m_{np}}{2 T(1+\cosh(m_{np}/T))}\mathrm{exp} \bigg[-\int^\infty_a \frac{\rmd a'}{a'} \tilde{\lambda}_{n\to p} \bigg]\\
			&\times\bigg\{-\tilde{\lambda}_{n\to p}\frac{\Delta m_{np}}{m_{np}}+\int^\infty_a\frac{\rmd a'}{a'}\tilde{\lambda}_{n\to p}\bigg[\frac{m_{np}X^{eq}_n}{T}\frac{\Delta m_{np}}{m_{np}}\\
			&\qquad\qquad-\frac{6 {g_A}_n^2}{1+3{g_A}_n^2}\frac{\Delta {g_A}_n}{{g_A}_n}-2\frac{\Delta G_F}{G_F}-\frac{m_{np} J'}{TJ}\frac{\Delta m_{np}}{m_{np}}\\
			&\qquad\qquad+\frac{3\zeta(3)}{2J}\frac{m^2_e}{T^2}\left(\frac{\Delta m_{e}}{m_{e}}-\frac{\Delta m_{np}}{m_{np}}\right)\bigg]\bigg\}\, .\label{eq:long_DXnW}
		\end{aligned}
	\end{equation}
\end{mdframed}
\subsubsection{Variation in the neutron abundance at BBN}
After weak freeze-out, the neutron abundance is modified by neutron decay. To find the variation in the neutron abundance at BBN, we vary equation \ref{eq:xbbn} to find 
\begin{align}
	\frac{\Delta X_{n,BBN}}{X_{n,BBN}}&=\frac{\Delta X_{n,W}}{X_{n,W}}-\bigg(\int^{a_{BBN}}_{a_W}\frac{\rmd a}{a H}\Delta\Gamma_n\bigg)-\frac{\Gamma_n}{H}\frac{\Delta a}{a}\Bigg|^{a_{BBN}}_{a_{W}}\\
	&\approx\frac{\Delta X_{n,W}}{X_{n,W}}-\bigg(\int^{a_{BBN}}_{a_W}\frac{\rmd a}{a H}\Delta\Gamma_n\bigg)+\frac{\Gamma_n}{H}\Bigg|_{a_{BBN}}\frac{\Delta T_{BBN}}{T_{BBN}}\label{eq:DXnBBN}\, ,
\end{align}
where we dropped the $\frac{\Delta T_W}{T_W}$ term because at weak freeze-out  $\frac{\Gamma_n}{H}\ll1$.
The variation in the neutron lifetime is given by 
\begin{equation}
	\frac{\Delta\Gamma_n}{\Gamma_n}=\frac{6 {g_A}_n^2}{1+3 {g_A}_n}\frac{\Delta {g_A}_n}{{g_A}_n}+2\frac{\Delta G_F}{G_F}+5\frac{\Delta m_e}{m_e}+\frac{m_{np}P'}{m_{e}P}\left(\frac{\Delta m_{np}}{m_{np}}-\frac{\Delta m_e}{m_e}\right)\, . \label{eq:DGamman}
\end{equation}\\

\subsubsection{Variation in the helium abundance}
The variation of equation \ref{eq:XHe} gives
\begin{align}
	\frac{\Delta Y_p}{Y_p}&=\frac{\Delta X_{n,BBN}}{X_{n,BBN}}\,. \label{eq:DYhe}
\end{align}
The remaining piece of our calculation is to determine $\frac{\Delta T_{BBN}}{T_{BBN}}$.
Recall that the time of BBN is determined by the deuterium abundance reaching a critical value allowing the efficient fusing of deuterium into \He. Using equation \ref{eq:saha}, we can find $T_{BBN}$ by examining 
\begin{equation}\label{eq:saha_at_BBN}
	X_D^\mathrm{crit}=\frac{24\zeta(3)}{\sqrt{\pi}}\eta_b (1-X_{n,BBN})X_{n,BBN} \left(\frac{T_{BBN}}{m_p}\right)^{3/2}e^{B_D/T_{BBN}}\, .
\end{equation}
Note that the left-hand side is exponentially dependent on $B_D/T_{BBN}$. Therefore, any shift in $B_D$ must be compensated by a shift in $T_{BBN}$, indicating that $T_{BBN}$ depends mostly on $B_D$, i.e.
\begin{equation}
	\frac{\Delta T_{BBN}}{T_{BBN}}\approx\frac{\Delta B_D}{B_D}\, .
\end{equation} 
Combining everything, we get 
\begin{mdframed}
	\begin{equation}
		\begin{aligned}
			\frac{\Delta Y_p}{Y_p}&=\frac{\Delta X_{n,W}}{X_{n,W}}+\frac{\Gamma_n}{H}\frac{\Delta B_D}{B_D}\Bigg|_{a_{BBN}}-\int^{a_{BBN}}_{a_W}\frac{\rmd a}{a}\frac{\Gamma_n}{H}\bigg(\frac{6 {g_A}_n^2}{1+3 {g_A}_n}\frac{\Delta {g_A}_n}{{g_A}_n}\\
			&\quad+2\frac{\Delta G_F}{G_F}+5\frac{\Delta m_e}{m_e}+\frac{m_{np}P'}{m_{e}P}\left(\frac{\Delta m_{np}}{m_{np}}-\frac{\Delta m_e}{m_e}\right)\bigg)\label{eq:longDYhe}\, .
		\end{aligned}
	\end{equation}
\end{mdframed}
Using equations \ref{eq:deltame}, \ref{eq:deltamnp}, and \ref{eq:deltaBD} for the variations in the fundamental constants we can now find the change in the \He abundance in terms of the field value of $\phi$ as a function of redshift. $\phi$ as a function of redshift was found using the evolution as described in section \ref{sec:DMevo}. We place constraints on the DM couplings by requiring that the resulting shift in the \He abundance keeps the \He abundance in the 95\% confidence interval of the observed abundance, {\it i.e.} we require that $Y^{ex}_p-Y^{th}_p-2 \sigma_{Y_p}<\Delta Y_p <Y^{ex}_p-Y^{th}_p+2 \sigma_{Y_p}$ with $\sigma_{Y_p}$ being the uncertainty of the theory prediction and the experimental uncertainty added in quadrature.  


\subsection{Results}
\label{sec:results}

\begin{figure}[b]
	\centering
	\includegraphics[width=1.0\textwidth]{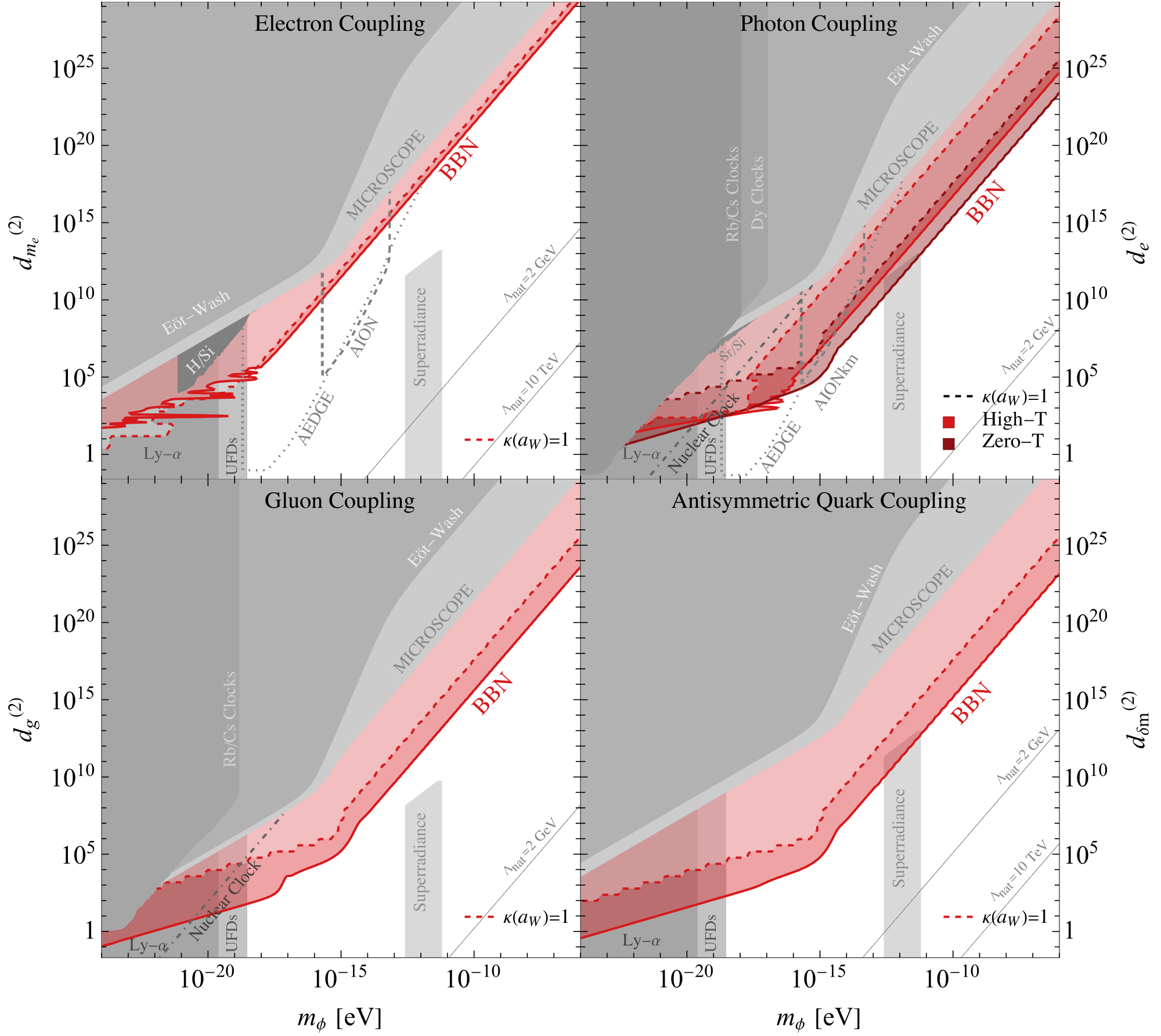}
	\caption{Constraints on quadratic couplings of ULDM to electrons, quarks, gluons, and photons.  Red regions indicate the BBN constraints derived in this paper. Other constraints are given in shades of gray. On the photon coupling plot, the conservative high-temperature approximation of the thermal mass yields the constraint in lighter red while the optimistic zero-temperature (i.e., without backreaction) result yields the constraint in darker red, see section \ref{sec:photonmass} for details.   The dashed red lines indicate where the approximation used in this paper breaks down and a more detailed analysis is needed.  The grey regions show parameter space ruled out by prior work from the E\"{o}t-Wash~\cite{EotWash_EP_2008} and MICROSCOPE~\cite{MICROSCOPE_2019} experiments, atomic clock experiments~\cite{SYRTE2016,Hees_2018} as well as constraints from Lyman-$ \alpha $~\cite{Rogers:2020ltq}, UFDs~\cite{Dalal:2022rmp} and superradiance~\cite{Baryakhtar:2020gao,Davoudiasl:2019nlo}. Also shown, with dashed gray lines, are projected constraints from the AION and AEDGE experiments~\cite{Badurina_2020,El_Neaj_2020} as well as projections for nuclear clocks~\cite{Antypas:2022asj}. Following the discussion in section \ref{sec:model}, we show the line above which the low values of $ m_\phi $ may be rendered unnatural by large loop corrections, assuming a cutoff of $ \Lambda = 10 $ TeV and $ \Lambda = 2 $ GeV. For the gluon and photon plots, the line for a $ \Lambda = 10 $ TeV cutoff lies outside the range shown.}
	\label{fig:results}
\end{figure}

The constraints on quadratically-coupled ULDM obtained from BBN in this work are shown as solid red lines in figure \ref{fig:results}. For masses larger than about $10^{-14}\mathrm{~eV}$, the evolution is in the bare mass dominated regime both during and after weak freeze-out. For these masses, the constraints scale as $d^{(2)}_i\propto m_\phi^{2}$ regardless of coupling. For the quark and gluon couplings, the field is Hubble frozen before BBN for masses lower than $10^{-19}\mathrm{eV}$. This leads the constraint to scale as $d^{(2)}_i\propto m_\phi^{1/2}$ in this low-mass regime. At intermediate masses, the field is Hubble frozen-in during or between weak freeze-out and BBN and the behavior is more complicated. For the electron coupling, the thermal mass is relevant for the evolution of the dark matter for masses less than  $10^{-18}\mathrm{eV}$. This leads to slow oscillations on the timescale of relevant BBN dynamics and the complex behavior shown in figure  \ref{fig:results}. For the photon coupling, we show two different constraints for two different estimates of the effective mass. The dark red line neglects the thermal mass of $\phi$ and the red line uses the thermal mass calculated in the high-temperature limit. We expect that the true constraint, calculated using the exact thermal mass, would be in between these two bounds as explained in section \ref{sec:photonmass}.

The BBN constraints derived in this work improve on the constraints derived in \cite{Stadnik_2015} by including the effects of the thermal mass on the evolution of the dark matter and by treating weak freeze-out using the full kinetic description. For the electron coupling, at masses of at least $m_\phi \gtrsim 10^{-14}\mathrm{eV}$ the two effects lead to a constraint about two orders of magnitude weaker than the constraint given in \cite{Stadnik_2015}. This reduced constraint is explained by the thermal mass which reduces the field value at the BBN era. For the quark coupling, the thermal mass is insignificant and the constraint on this coupling is instead enhanced by about a factor 2 relative to \cite{Stadnik_2015} in the $ m_\phi \gtrsim 10^{-14}\mathrm{eV}$ regime.
In all cases, for masses below $10^{-14}\mathrm{eV}$ we see nontrivial behavior not seen in  \cite{Stadnik_2015}. For the photon coupling, our results indicate only the region of parameter space in which a full calculation of the thermal effects is expected to place the constraint. The constraint on the photon coupling given in \cite{Stadnik_2015} lies within the region where we expect the constraint to lie. Therefore, we cannot show a change in the photon coupling constraint without a more detailed evaluation of thermal effects, which we leave to future work. The gluon coupling was not treated in \cite{Stadnik_2015} and thus the constraints on the gluon coupling derived in this work are completely novel.  

Our analysis is valid in the regime where $d^{(2)}_i\varphi^2\ll 1$. In this regime, we can neglect the higher-order interactions of $\phi$ with the SM that would be present in a UV completion and treat $\phi$ as a small perturbation to the SM. Importantly, we can neglect the effect of the thermal quartic and other higher-order terms on the evolution of the ULDM. Because the amplitude of $\phi$ decreases with redshift, the threshold $d^{(2)}_i\varphi^2\sim 1$ could have been crossed sometime in the early Universe; if this happens after weak freeze-out then details of the UV completion are needed to determine the impact of $\phi$ on BBN. The parameter space in which the condition that $ d^{(2)}_i\varphi^2\ll 1 $ is violated at any time after weak freeze-out corresponds to the lightly shaded regions above the dashed red lines in figure \ref{fig:results}. In these regions, a model-dependent treatment is required for any given UV completion.


\section{Other Constraints}
\label{sec:constraints}
Scalar ULDM with the couplings discussed in this paper is subject to several constraints in addition to the BBN bound derived here.  These include constraints from searches for weak equivalence principle (WEP) violation, atomic clock experiments directly looking for varying fundamental constants, probes of cosmological structure formation, and astrophysical constraints. 
\subsection{Weak Equivalence Principle Violation Searches}
Searches for violation of the WEP look for accelerations on test masses in addition to those expected from gravity. Such accelerations can violate the universality of free fall (UFF) and can be generated by long-range Yukawa forces mediated by $ \phi $.
Classically, this effect appears because the $ \phi $-SM couplings perturb fundamental constants which leads to perturbations of the binding energy and thus the mass of SM matter.
In general, the $\phi$-dependence in SM matter will depend on the material composition, so that the resultant force will not respect the WEP. 
In the presence of background $ \phi $-field gradients or oscillations, this $ \phi $-dependence in the mass of SM matter lead to apparent WEP violations that are highly constrained by experiment.


Such constraints on apparent WEP violation depend sensitively on the background field configuration, which was studied in \cite{Hees_2018}. For linear interactions, the $ \phi $-SM interactions generate a static field configuration around the Earth itself, the gradient of which leads to WEP violation constraints. For quadratic interactions, no static solution beyond the trivial one exists. Instead, $\phi$ gets an effective mass inside the Earth and other massive bodies (see e.g.~\cite{Stadnik:2020bfk}). This effect screens the $\phi$ field\footnote{If the $ \phi -$SM coupling appears with a negative sign, then an enhancement of the field value can instead be found. We do not consider such scenarios here. See, e.g., discussions in \cite{Sibiryakov_2020,Hees_2018}.},  creating gradients in the $\phi$ field in the vicinity of the earth.

The strongest constraints from WEP violation searches are set by the E\"{o}t-Wash experiment~\cite{EotWash_EP_2008} and the MICROSCOPE experiment~\cite{MICROSCOPE_2019}.
The E\"{o}t-Wash experiment searches for equivalence principle violation by monitoring a torsion balance with masses made of dissimilar materials (beryllium and titanium) and looking for torques generated by differing force per mass on the test bodies. For quadratic interactions, $\phi$ acquires an induced mass from the interaction with the  Earth, which screens $ \phi $. This effect limits the sensitivity of Earth-based experiments. The MICROSCOPE experiment is a space-based experiment that looks for differing gravitational accelerations on test masses in orbit. Because of the reduced screening, the MICROSCOPE experiment provides stronger constraints than the E\"{o}t-Wash experiment across all masses for quadratic interactions. These constraints are shown in figure \ref{fig:results}.


This discussion of WEP violation constraints follows the derivation in \cite{Hees_2018}. In this work, it is assumed that far away from the Earth $\varphi$ approaches 
\begin{equation}
	\varphi\simeq\sqrt{\frac{\rho_{\rm DM}}{2\pi M_{\rm pl}^2 m_\phi^2}}\cos(m_\phi t)
\end{equation}
with $\rho_{\rm DM}$ the local DM density. However, this assumption neglects the virial velocity of the DM; the virial velocity can be neglected when the wavelength of the DM is larger than the radius of the Earth. However, for $m_\phi \gtrsim 10^{-11}$ the wavelength is comparable to or smaller than the size of the Earth and the WEP constraints derived in \cite{Hees_2018} and displayed in Fig.~\ref{fig:results} may need to be modified. Note that this is the first time that the quark and gluon constraints from WEP searches have been presented for quadratic couplings.

\subsection{Experiments looking for  Varying Fundamental Constants}
Precision measurements can search for temporal variation in fundamental constants in labs on Earth and in space. By examining the stability of different atomic frequency standards relative to each other, tight constraints can be put on the temporal variation in the ratios of different atomic transitions (not necessarily of the same species). This puts strong constraints on the coupling of ULDM to photons, quarks and gluons \cite{Clocks2015,Hees_2018}. The most sensitive clock experiments are not sensitive to the electron coupling due to approximate cancellations for the frequency comparisons they make. Future improvements in this field may come from the development of nuclear clocks based on the nuclear transition between the ground state and the first excited state of the $ ^{229} $Th isotope. If such a clock is realized, the current constraints could be improved by many orders of magnitude. For a review of recent developments in the field, e.g., see \cite{Antypas:2022asj}. 

Matter-wave interferometry experiments are very sensitive to potential time variation in atomic transition energy~\cite{MattWave_2018}. A large number of upcoming experiments will probe interesting parameter space. A recent summary of progress in this field can be found in \cite{Antypas:2022asj}. Among the most sensitive of such atom interferometers are the Earth-based AION~\cite{Badurina_2020} and the space-based AEDGE~\cite{El_Neaj_2020}. The  Earth-based experiment is hampered by the screening effects discussed in the previous section. These experiments are also sensitive to the electron and photon couplings. 
\subsection{Structure Formation}
Cosmological structures cannot form on length scales smaller than the de Broglie wavelength of the DM. To correctly reproduce structure formation, the DM must therefore be sufficiently heavy. Observations of dwarf Milky Way satellites require $m_\phi\gtrsim 3\times 10^{-21}~\mathrm{eV}$ \cite{Lague:2021frh, Dwarf2020,DES:2020fxi}. Similar constraints come from measurements of the subhalo mass function and observations of stellar streams \cite{Subhalo2020,Banik:2019smi}.  The large de Broglie wavelength also delays structure formation compared to standard cosmology, so complementary constraints arise from observations of small-scale structure at high redshift. For example, the Lyman-$\alpha$ forest flux power spectrum requires $m_\phi\gtrsim 2 \times 10^{-20}~\mathrm{eV}$ \cite{Rogers:2020ltq}. 

The strongest lower bound on the mass of ULDM comes from observations of ultra-faint dwarf galaxies (UFDs). The wave-like properties of ULDM cause DM density fluctuations that transfer energy to stars through gravitational interactions leading to dynamical heating of dwarf galaxies. Measurements of the velocity dispersion of the UFDs Segue 1 and Segue 2 require $m_\phi\gtrsim 3\times10^{-19} \mathrm{eV}$~\cite{Dalal:2022rmp}. Similar constraints were claimed from observation of the dwarf galaxy Eridanus II~\cite{Eridanus2019}. However, that analysis made strong assumptions about the solitonic core of the DM halo of Eridanus II, which were relaxed in a more detailed analysis~\cite{Eridanus_2021}.


\subsection{Astrophysics}
Astrophysical probes are also sensitive to the presence of interacting ULDM. In particular, pulsar timing arrays can be used to constrain very light ULDM in a manner complementary to the structure formation arguments discussed above~\cite{Blas:2016ddr,Kaplan:2022lmz}. These constraints are weaker than the constraints from structure formation and are thus not shown here. 
At higher masses, constraints arise from superradiance.
Bosonic radiation incident on a rotating black hole can be amplified through a process called black hole superradiance in which energy and angular momentum are extracted from the black hole in a manner similar to the Penrose process~\cite{teukolsky1974perturbations} and results in the spin-down of black holes~\cite{Brito_2020}.  
The observation of old, rapidly rotating black holes can therefore be used to rule out the existence of light bosons
\cite{Arvanitaki:2009fg,Arvanitaki:2010sy,Arvanitaki:2014wva,Brito:2015oca,Cardoso:2018tly,Stott:2018opm,Stott:2020gjj,Mehta:2020kwu}.  Observations of solar mass black holes can be used to constrain light scalars with masses in the interval $[2.7\times10^{-13},6.1\times 10^{-12}]~\mathrm{eV}$ \cite{Baryakhtar:2020gao}. Supermassive black holes can also be used to constrain DM mass in the interval $[2.9\times10^{-21},1.6\times 10^{-17}]~\mathrm{eV}$ ~\cite{Unal:2020jiy,Arvanitaki:2014wva,Stott:2018opm,Davoudiasl:2019nlo}, although the spin measurements are not as robust as for solar mass black holes and less is understood about the accretion disks, whose properties may disrupt the superradiance process~\cite{Brito_2020,Du:2022trq}.
Quadratically coupled ULDM will get an effective mass from interaction with the accretion disk in the vicinity of a black hole, potentially modifying the dynamics of superradiance. In the absence of a detailed analysis of this effect, we restrict the superradiance constraints to couplings where the induced mass from the accretion disk is subdominant to the bare mass,
\begin{equation}
	m_{\rm induced}^2=d_i^{(2)}\frac{2\pi}{M_{\rm Pl}^2}2Q_i\rho_{\rm BH}\ll m_\phi^2\, ,
\end{equation}
where $\rho_{\rm BH}$ is the density of the accretion disc at the radius of the boson cloud, and $Q_i$ is the dilatonic charge of the accretion disk as in \cite{Hees_2018}.
Furthermore, we neglect any self-interactions. Such self-interactions can disrupt superradiance, potentially weakening the bound further~\cite{Fukuda:2019ewf,Mathur:2020aqv,Baryakhtar:2020gao}.    


\section{Conclusions}
\label{sec:conclusions}

In this work, we examined the effect of ultralight scalar DM with quadratic couplings on the predicted helium abundance produced by Big Bang nucleosynthesis.
Figure \ref{fig:results} shows the constraints derived in this work. In addition, we also show constraints from the E\"{o}t-Wash and MICROSCOPE experiments, atomic clocks experiments, and projected constraints from the AION and AEDGE matter-wave interferometry experiments. This is the first time that the quark and gluon constraints from WEP searches have been presented for quadratic couplings. 
In this work, we treated weak freezeout in the full kinetic description and accounted for backreaction from the SM on the evolution of the DM. From the impact of ULDM on BBN, we constrain the couplings of ULDM to electrons, quarks, and photons for DM masses ranging from about $10^{-19}\mathrm{eV}$ to $10^{-4}\mathrm{eV}$. This updates the result in \cite{Stadnik_2015} that treated weak freezeout in the instantaneous approximation and neglected backreaction from the SM on the evolution of the DM. 
For a significant range of parameter space, we show that the bounds from BBN are the strongest constraints on quadratically coupled ultra-light dark matter. We show that these BBN bounds are significantly modified by the effects of the backreaction and the full kinematic decoupling studied in this work.

\section*{Acknowledgments}
We thank Junwu Huang and Masha Baryakhtar for helpful comments on an earlier version of this draft and Andreas Ekstedt for useful discussions. This work is supported in part by NSF CAREER grant PHY-1944826, in part by the Deutsche Forschungsgemeinschaft under Germany’s Excellence Strategy - EXC 2121 “Quantum Universe” - 390833306, and in part by the project “CPV-Axion” under the Supporting TAlent in ReSearch@University of Padova (STARS@UNIPD), and by the INFN Iniziative Specifica APINE.
\bibliographystyle{JHEP}
\bibliography{lib.bib}

%
%

\end{document}